# Harnessing the Potential of Volatility: Advancing GDP Prediction

*Ali Lashgari* *

**ABSTRACT**

This paper presents a novel machine learning approach to GDP prediction that incorporates volatility as a model weight. The proposed method is specifically designed to identify and select the most relevant macroeconomic variables for accurate GDP prediction, while taking into account unexpected shocks or events that may impact the economy. The proposed method's effectiveness is tested on real-world data and compared to previous techniques used for GDP forecasting, such as Lasso and Adaptive Lasso. The findings show that the Volatility-weighted Lasso method outperforms other methods in terms of accuracy and robustness, providing policymakers and analysts with a valuable tool for making informed decisions in a rapidly changing economic environment. This study demonstrates how data-driven approaches can help us better understand economic fluctuations and support more effective economic policymaking.

* Ph. D. Candidate; Kansas State University, Manhattan, KS, USA email: alilashgari@ksu.edu
Thanks to Dr. Bachmeier and Dr. Lee

**INTRODUCTION**

Forecasting is a valuable tool in various academic fields, including geography, business, mathematics, physics, and medicine. It assists geographers in predicting changes in weather patterns, natural disasters, and land use, leading to improved urban planning and disaster preparedness (Mohammadpouri, Saeid et al. , 2023 , Kazemi Garajeh, M. et al., 2023, He, Feng et al., 2023) Additionally, it can be used to forecast energy consumption and demand in urban areas, providing insights for designing appropriate lighting plans and transportation energy strategies (Kazemidemneh, M. et al.,2018, Arjomandnia, R. et al., 2023).

Forecasting plays a critical role in mathematics and physics by helping understand and predict complex phenomena. In physics, it is employed to model and anticipate the behavior of physical systems, including quantum mechanics and fluid dynamics. For instance, researchers have developed gas-filled photonic microcells by modifying the end of photonic bandgap fibers, resulting in high transmission efficiency and moderate line center accuracy. These cells can be interconnected, albeit with a slight decrease in efficiency. This advancement holds promise for monitoring and predicting changes in the performance of photonic microcells over time, potentially enhancing the efficiency and accuracy of laser micro/nano-machining processes. Physicians can leverage forecasting to improve the prevention and management of conditions like PCOS and T2DM. By predicting an individual's risk of developing these conditions based on lifestyle, genetics, and other relevant factors, forecasting enables the creation of personalized prevention and management plans. For instance, if someone has a higher risk of developing T2DM, their doctor may recommend specific lifestyle changes and/or medications to delay or prevent the onset of the disease. Likewise, forecasting can aid in predicting a patient's response to different treatments for PCOS, assisting doctors in tailoring the treatment plan accordingly (Xue, Lianjie, et al., 2021, Hosseini-Zavareh, Sajed, et al.,2019, Luder, Ryan J., et al. 2016, Hosseini-Zavareh, Sajed, et al., 2018, Shoaibinobarian, et al., 2022, Jafari, Naser, et al., 2023).

The rapid growth of big data and artificial intelligence (AI) has posed challenges for forecasting financial time series data. Traditional methods relying on statistical assumptions struggle to handle the abundance of complex correlations present in financial data. As a result, researchers have shifted their focus towards nonparametric methods and machine learning (ML) algorithms. ML algorithms can automatically learn patterns and relationships from large datasets, making accurate predictions without explicit programming. They excel at processing vast amounts of financial data, including unstructured sources like news articles and social media sentiment. AI and ML models adapt to changing market conditions, continuously updating forecasts and capturing complex nonlinear relationships. These techniques are widely used in financial forecasting tasks such as predicting stock prices, market trends, risk assessment, and fraud detection. However, challenges like high-quality data requirements, interpretability, and avoiding overfitting or underfitting must be addressed to ensure reliable and robust forecasts. The combination of big data and AI has transformed financial time series forecasting. AI and ML techniques enable the analysis of extensive financial data, adapt to evolving market dynamics, and incorporate unstructured information sources. These methods provide advantages such as quick processing, capturing complex patterns, and handling diverse data types. AI and ML models find applications in predicting stock prices, optimizing portfolios, assessing risks, and detecting fraudulent activities. However, challenges like data quality, biases, model interpretability, and avoiding overfitting remain significant considerations. With further advancements, AI and ML are poised to continue playing a pivotal role in enhancing forecasting accuracy and supporting decision-making in the financial industry (Nematira, 2022, Lashgari, 2022).

Accurate GDP forecasting is critical for policymakers, investors, and economic analysts assessing an economy's overall health and performance. Traditional regression algorithms frequently fail to capture the dynamic nature of economic data, resulting in unsatisfactory forecasts. When using regression models with more explanatory variables than observations, economists face a substantial hurdle. Such scenarios frequently provide incorrect results and overfitting, prompting variable reduction or the gathering of further data.

To address this issue, numerous traditional variable selection procedures, including as forward selection, backward elimination, stepwise selection, are often used. These approaches seek to identify the most influential factors inside regression models. Traditional procedures are used to evaluate significant variables by examining their statistical significance and overall model goodness-of-fit using p-values, adjusted R-squared values, and AIC/BIC values. By balancing model complexity and model fit, these strategies attempt to discover significant factors that explain the outcome while avoiding overfitting (Draper, N. R. & Smith, H 1998).

Principal Component Analysis (PCA) is another useful technique in machine learning for reducing dimensionality while keeping important information from the original data. Karl Pearson invented it in 1901, and

Harold Hotelling expanded on it in the 1930s. The goal of PCA is to convert correlated variables into uncorrelated principal components, capturing essential patterns and relationships in the dataset (Jolliffe, I. T. 2002).

Regularization methods used in regression models with several predictor variables include ridge regression, LASSO, and Elastic Net. Ridge regression (Hoerl, A. E. & Kennard, R. W. 1970) addresses collinearity by introducing a penalty term, whereas LASSO (Tibshirani, R. 1996) encourages sparsity by removing irrelevant variables. Elastic Net (Zou, H., & Hastie, T. 2005) uses both strategies to manage strong predictor correlation. These approaches, however, can produce biased results and fail to discover truly relevant variables.

This problem is also addressed by the Adaptive Lasso method, which adjusts the penalty parameter based on the relevance of predictor factors. It weights the penalty by the inverse of the absolute value of the calculated coefficients, making it easier to identify relevant factors. Adaptive Lasso improves variable selection and produces accurate predictions by punishing less significant factors more harshly.

To improve forecast accuracy in time series models, the Lag Weighted Lasso approach was introduced. It takes into account both coefficient size and lag effects by penalizing each coefficient differently. In forecasting, the Lag Weighted Lasso outperforms the standard LASSO and Adaptive Lasso, especially for time series models with no seasonality and diminishing variable effects with increasing lag length.

(Park & Sakaori 2013) proposed a new method called the lag weighted lasso for time series models. While the adaptive lasso can consistently identify the true model in regression models, it fails to account for lag effects, which are essential for a time series model. The lag weighted lasso improves the forecast accuracy of a time series model by imposing different penalties on each coefficient based on weights that reflect not only the coefficient size but also the lag effects. The paper demonstrates that the lag weighted lasso outperforms the lasso and the adaptive lasso in forecast accuracy, especially for time series models with no seasonality, when the variable effects decrease with increasing lag length.

This paper introduces the Volatility Weighted Lasso approach, a novel method aimed at enhancing forecast accuracy in time series models. The study focuses on applying this approach to forecast GDP in the United States, utilizing macroeconomic variables as estimators. Additionally, the SHAP approach (Lundberg, S. M., & Lee, S.-I. 2017) is employed to assess the relevance of each variable.

The methodology section provides a comprehensive overview of the Volatility Weighted Lasso approach. In the simulation section, a comparative analysis is presented, evaluating the effectiveness of volatility for the proposed method. Then the real data of macroeconomics variables show how the accuracy improved by volatility. In addition, the significance of each variable are provided through the utilization of the SHAP method.

**METHODOLOGY**

Consider the multiple linear regression model for time series data is:

$$y_t = \beta_0 + \beta_1 x_{1,t} + \beta_2 x_{2,t} + \cdots + \beta_k x_{k,t} + \epsilon_t \quad (1)$$

$$y = X\beta + \varepsilon \quad (2)$$

Equation (2) is a vector form where y is a response vector, X is a vector of variables, $\beta$ is a coefficient vector, and $\varepsilon$ is an error vector. The Least Absolute Shrinkage and Selection Operator (Lasso) considers both variable selection and regularization to enhance the prediction accuracy and interpretability of the statistical model. Introduced by Robert Tibshirani in 1996, it is a type of linear regression that uses a shrinkage constraint on the coefficients, which leads to some coefficients being exactly zero. This means that the model is less likely to overfit, can handle collinearity in input variables, and implicitly does feature selection. It selects a subset of the provided covariates for use in the final model.

$$\hat{\beta} = \arg\min \left( y - \sum_{j=1}^{p} \beta_j x_j \right)^2 + \lambda \sum_{j=1}^{p} \hat{\omega}_j |\beta_j| \quad (3)$$

The adaptive Lasso introduced by Zou in (2006) estimates coefficients by equation (3), where $\hat{\omega}_j$ is a set of weights such that $|\beta_j|^\gamma, \gamma > 0$. It is an extension of the Lasso where the penalty term is weighted by an initial estimate of the inverse absolute coefficients. It adds an adaptive weight to each coefficient in the Lasso penalty, which can lead to consistent parameter estimation (it can recover the true coefficients) and is claimed to have the "oracle property" under certain conditions (Fan J, Li R 2001).

Adaptive Lasso is a modification of the LASSO method, adds a penalty term to the regression model to encourage sparsity and select important variables. However, the Adaptive Lasso method differs from LASSO in how the penalty term is constructed. In LASSO, a fixed penalty parameter is used to shrink the coefficients of the predictor variables towards zero. Lasso provides sparse solutions that are biased, so the variables that lasso selects as meaningful can differ from the truly meaningful variables. The problem with this approach is that it does not take into account the different degrees of importance of the predictor variables. In other words, some variables may have a greater impact on the outcome than others, and they should be penalized differently.

An estimator is oracle proposed by Fan and Li (2001) if it can correctly select the nonzero coefficients in a model with probability converging to one, and if the nonzero coefficients are asymptotically normally distributed.

Adaptive Lasso solves this problem by adapting the penalty parameter to the importance of the predictor variables. Specifically, the penalty parameter is weighted by the inverse of the absolute value of the estimated coefficients in the initial stage of the regression analysis. This means that variables with larger coefficients are less penalized than those with smaller coefficients, and this leads to a better identification of the important variables. If a variable is important, it should have a small weight. This way it is lightly penalized and remains in the model. If it is not important, by using a large weight we ensure that we get rid of it and send it to zero.

In the new approach, the new weight is a function of volatility, where $\sigma_{jt}^2$ will be obtained from GARCH (1,1) model. The idea behind this model is the small changes are not believed to be useful as a predictor in the macroeconomics model, and the large changes are likely to have a large effect on the economy.

**SIMULATION**

To compare the volatility weighted lasso with the lasso and the adaptive lasso for the time series model, simulation studies has been examined. In this simulation, the objective is to understand how these methods handle the challenges posed by time series data, such as serial dependence and changing volatility. To generate the data, the number of observations, n, is 100, and the number of predictors, p, is 5. The predictor variables, X, are generated as autoregressive processes of order 1, with each predictor following its unique AR (1) process. The error term, $\varepsilon$, that also follows an AR (1) process but with a time-varying volatility. The volatility is determined by a sinusoidal function that ranges from 1 to $2\pi$. The Lasso Regression (LASSO) approach applies regularization to shrink irrelevant coefficients towards zero, without considering the volatility in the error term. The Adaptive Lasso method (AD LASSO) incorporates a preliminary fit using LassoLarsIC to determine weights based on the coefficients' absolute values, which are then used to adjust the penalization in the subsequent regression model. The Volatility Weighted Lasso method (VW LASSO) incorporates GARCH modeling to calculate volatility weights, capturing the changing volatility in the error term. These weights are then incorporated into the penalization term, allowing for increased penalty during high volatility periods. The estimated coefficients obtained from each method highlight the impact of volatility on coefficient estimation. (Table I)

**Table I.** Simulation results

|  | Ceof 1 | Ceof 2 | Ceof 3 | Ceof 4 | Ceof 5 |
|---|---|---|---|---|---|
| **LASSO** | 3.069 | -1.78 | 0 | 0 | 0 |
| **AD LASSO** | 2.78 | -5.406 | 7.37777265e-18 | 6.00993626e-18 | -6.201964e-18 |
| **VW LASSO** | 2.088 | -1.183 | 0 | 0.322 | -0.010 |

Through the simulation, the role of volatility in coefficient estimation is particularly evident in the Volatility Weighted Lasso method. High volatility periods can lead to larger residuals in the model. If the model does not account for this, it may incorrectly attribute these large residuals to the predictor variables. As a result, the Volatility Weighted Lasso method may assign non-zero coefficients to predictors during high volatility periods, even if they are not truly influential.

**REAL WORLD DATA**

To evaluate the practicality of the proposed method in relation to existing approaches, a real-world data is presented to demonstrate its effectiveness. The data includes quarterly GDP in the United States from Q1 1986 to Q4 2022. Variable screening by regularization methods shows which macroeconomics variables should be in the model(Firoozabadi, 2019, Goharipour, 2022). There are ten macroeconomics variables using to predict GDP.

- Inflation: The rate at which prices for goods and services are rising. (INFLATION)
- Housing starts: The number of new housing units that are under construction. (HOUSING)
- Interest rates: The cost of borrowing money. (PC)
- Business confidence: The degree of optimism or pessimism that business owners have about the economy (PCE)
- Government spending: The amount of money that the government spends on goods and services. government consumption expenditure. (GCE)
- Net exports: The difference between exports and imports. (NET)
- Employment levels: The number of people employed in the economy. (UNEMPLOY)
- Money supply: The amount of money circulating in the economy. (M1)
- Stock market performance: The performance of the stock market, as measured by various indices. (NASDAQ)
- Oil price (WTI)

The data is separated into two subgroups for forecasting GDP: the training set and the test set. The training set is used to construct and train the predictive model, whereas the test set is used to assess the model's performance. In this specific case, the GDP has been predicted using 10 macroeconomic variables.

To obtain a comprehensive understanding of the models' performance, we vary the size of the training set by splitting the data three times: 70%, 80%, and 90% for training, respectively. For each training set size, we apply the predictive models and generate forecasts for GDP and then optimize the solution by methods of Noroozi, 2022, Abdidizaji, 2021, Hajipou, 2021, Nematirad, 2023, and Ansari, 2022. To assess the accuracy of the forecasts, we calculate two commonly used evaluation metrics: the Root Mean Squared Error (RMSE) and the Mean Absolute Error (MAE). Table II

**Table I.** Accuracy of different models

| Model/Criteria | | MAE | RMSE |
|---|---|---|---|
| **Multiple Linear Regression** | train = 70 | 152.20322 | 184.24388 |
| | train = 80 | 838.02230 | 1197.23821 |
| | train = 90 | 196.80460 | 263.30660 |

| | train = 70 | 270.91236 | 306.43466 |
| --- | --- | --- | --- |
| **Lasso** | train = 80 | 129.96519 | 162.24995 |
| | train = 90 | 202.34137 | 243.86772 |
| | train = 70 | 682.24205 | 1155.40999 |
| **Ridge** | train = 80 | 277.67070 | 364.52109 |
| | train = 90 | 857.59601 | 957.73570 |
| | train = 70 | 236.99972 | 333.75552 |
| **Adaptive Lasso** | train = 80 | 195.08969 | 251.47372 |
| | train = 90 | 212.16052 | 240.49616 |
| | train = 70 | 191.254 | 302.08 |
| **Volatility Weighted Lasso** | train = 80 | 168.22 | 247.376 |
| | train = 90 | 273.38 | 264.55 |

The performance of several models in forecasting GDP with three different training set sizes is shown in the table above. Among the models tested, the Volatility Weighted Lasso model achieved relatively low MAE and RMSE values on average across all three training set sizes. This implies that adding volatility to the adaptive Lasso model as a weight increases its accuracy in GDP forecasting.

In Figure I, the SHAP model is employed to interpret the importance of each macro variable in predicting GDP. The results indicate that the PCE variable exhibits the highest impact on forecasting GDP. This means that changes in PCE have the most significant influence on the model's predictions for GDP. By analyzing the SHAP values associated with each variable, we can gain insights into the relative importance and contribution of different macro variables in the GDP prediction process. The prominence of PCE underscores its crucial role in understanding and forecasting economic growth, highlighting the significance of business confidence patterns in shaping overall economic activity.

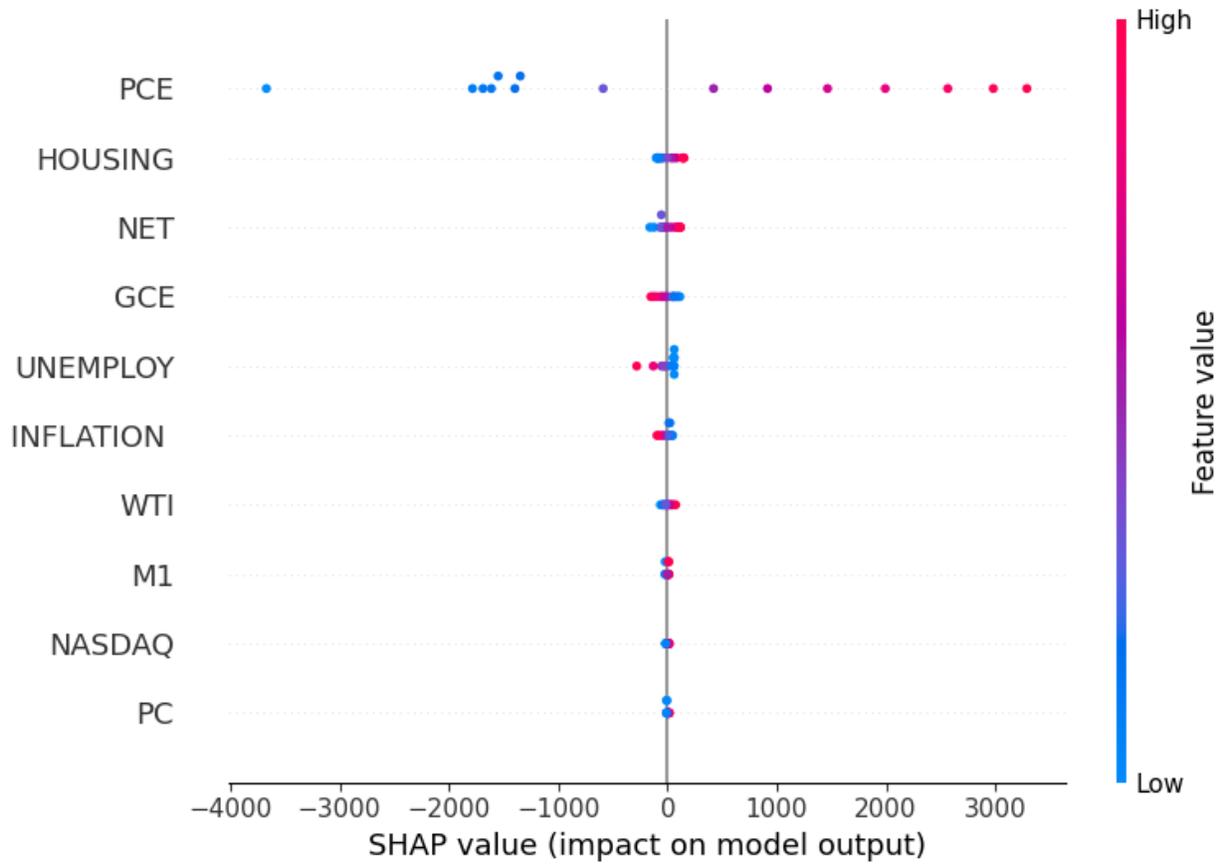

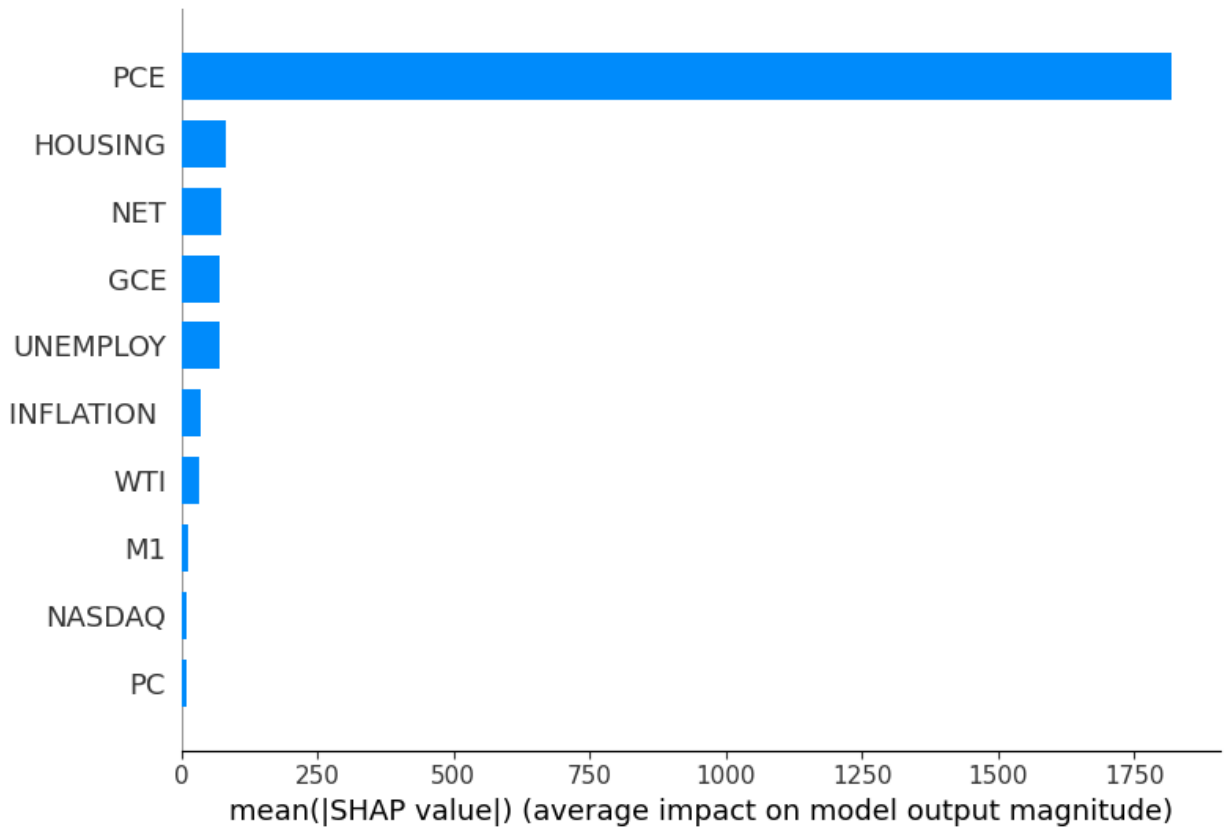

**Figure I.** SHAP Analysis

## CONCLUSION

In conclusion, this paper presents a novel machine learning approach, the Volatility-weighted Lasso, for GDP prediction that incorporates volatility as a model weight. The method is designed to identify relevant macroeconomic variables and account for unexpected shocks or events in economic forecasting. By comparing the proposed approach with existing techniques such as Lasso and Adaptive Lasso using real-world data, our findings demonstrate that the Volatility-weighted Lasso method outperforms other methods in terms of accuracy and robustness. The findings highlight the significance of volatility in forecasting GDP and emphasize its role as a crucial variable in economic modeling.

This study offers policymakers and analysts a valuable tool for making informed decisions in the face of a rapidly changing economic environment. The study underscores the potential of data-driven approaches in improving our understanding of economic fluctuations and supporting more effective economic policymaking. By incorporating volatility as a model weight, our approach provides a comprehensive framework for GDP prediction that can lead to better insights and more reliable forecasts. Future research can further explore the combination of the volatility weighted and lag weighted in economic forecasting domains and address any potential limitations or challenges associated with the approach.

**REFRENCES**